# Thermodynamic properties of the Superstatistics and Normal Statistics of the Schrodinger Equation with generalized trigonometric Pöschl–Teller potential.

C.O.Edet[1], P.O.Amadi[1], A.N.Ikot[1]  U. S. Okorie[1&2]  A. Taş[3], and G. Rampho[4].

[1]Department of Physics, Theoretical Physics Group, University of Port Harcourt, Choba, Nigeria.
[2]Department of Physics, Akwa Ibom State University, Ikot Akpaden, P.M.B. 1167, Uyo. Nigeria.
[3]Department of Physics, Mersin University, Mersin 33343, Turkey.
[4]Department of Physics, University of South Africa, South Africa.

## ABSTRACT

Analytical solutions of the Schrödinger equation for the generalized trigonometric Pöschl–Teller potential by using an appropriate approximation to the centrifugal term within the framework of the Functional Analysis Approach have been considered. Using the energy equation obtained, the vibrational partition function was calculated and other relevant thermodynamic properties. More so, we use the concept of the superstatistics to also evaluate the thermodynamics properties of the system. It is noted that the well-known normal statistics results are recovered in the absence of the deformation parameter $(q=0)$ and this is displayed graphically for the clarity of our results. We also obtain analytic forms for the energy eigenvalues and the bound state eigenfunction solutions are obtained in terms of the hypergeometric functions. The numerical energy spectra for different values of the principal $n$ and orbital $\ell$ quantum numbers are obtained. To show the accuracy of our results, we discuss some special cases by adjusting some potential parameters and also compute the numerical eigenvalue of the trigonometric Pöschl–Teller potential for comparison sake. However, it was found out that our results agree excellently with the results obtained via other methods

Correspondence email address: collinsokonedet@gmail.com

# 1. INTRODUCTION

Solving the radial Schrödinger equation is of great importance in nonrelativistic quantum mechanics, because it is well established that the wave function contains all the necessary information required to describe a considered quantum system [1-5]. It is well known that exact solutions of this equation are only possible for a few potential models, such as the Kratzer [6–7], Eckart potential [8-10], shifted Deng-Fan [11-14], Molecular Tietz potential [15-18] etc. The exact analytical solutions of the Schrödinger equation with some of these potentials is only possible for $\ell = 0$. For $\ell \neq 0$ states, one has to employ some approximations, such as the Pekeris approximation [17, 18], to deal with the orbit centrifugal term or solve numerically [19, 20].

Several mathematical approaches have been developed to solve differential equations arising from these considerations. They include the supersymmetric approach [21–24], ], Nikiforov–Uvarov method [25–27],asymptotic iteration method (AIM) [28-30], Feynman integral formalism [31–34], factorization formalism [35, 36], Formula Method[37] exact quantization rule method [38–43], proper quantization rule[44–48], Wave Function Ansatz Method[49] etc..

The trigonometric Pöschl-Teller potential was proposed by Pöschl and Teller [50] in 1933 and it has been used in describing diatomic molecular vibration. This potential can be written as

$$V(r) = V_1 \cos ec^2(\alpha r) + V_2 \sec^2(\alpha r) \tag{1a}$$

where parameters $V_1$ and $V_2$ describe the property of the potential well, whereas the parameter $\alpha$ is related to the range of this potential [51].

This potential has been applied to study diatomic molecular vibration. Ever since it was proposed in 1933, researchers have given much attention to the molecular potential. For example, Liu et al. [51] carried out Fermionic analysis with this potential.The bound state solutions have also been carried out in the relativistic regimeby Zhang and Wang [52],Chen [53], Candemir[54] and Hamzavi[55].

Very recently, Hamzavi and Rajabi [56] also studied the s-wave solutions of the Schrödinger equation for this potential using the Nikiforov–Uvarov method. Hamzavi and Ikhdair. [57] obtained the approximate solutions of the radial Schrodinger equation for the rotating trigonometric PT potential using the Nikiforov–Uvarov method. The energy eigenvalues and their corresponding eigenfunctions were calculated for arbitrary $\ell$-states in closed form.

Motivated by Ref.[50-57], we propose a modification to the trigonometric Pöschl-Teller potential. This we call the Generalized trigonometric Pöschl-Teller potential. This potential is given as;

$$V(r) = V_1 \cos ec^2(\alpha r) + V_2 \sec^2(\alpha r) + V_3 \tan^2(\alpha r) + V_4 \cot^2(\alpha r) \tag{1b}$$

where parameters $V_1$, $V_2$, $V_3$ and $V_4$ describe the property of the potential well, whereas the parameter $\alpha$ is related to the range of this potential.

The fundamental reason for studying the thermodynamics properties of a given system is to calculate its vibrational partition function. The partition function which explicitly depends on temperature, aids us to obtain other thermodynamics properties. The vibrational partition function for certain potential models can easily be obtained by calculating the rotation–vibrational energy levels of the system whose applications are widely used in statistical mechanics and molecular physics [58,59].Different mathematical approaches have been employed by many researchers in evaluating partition function such as Poisson summation fornula [60], commulant expansion method [61], standard method [62] and Wigner–Kirkwood formulation [63]. Superstatistics is one of the most important and attractive topics in statistical mechanics. Superstatistics is a superposition of different statistics: One given by ordinary Boltzmann factor and another given by the fluctuation of the intensive parameter such as the inverse temperature. Superstatistics describe non-equilibrium systems with a stationary state and intensive parameter fluctuations and contains Tsallis statistics as a special case[64-72].It is therefore the primary objective of the present work to study to solve the Schrodinger equation for non-zero angular momentum with the generalised trigonometric Posch-Teller using the Functional Analysis Approach. We will also use the resulting energy equation to find the partition function which will enable us to calculate other thermodynamics properties via stastical mechanics and superstatistics mechanics approach.

This paper is organized as follows. In Sect. 2 we derive the bound states of the Schrodinger equation with the generalised trigonometric PT potential using the FAA. In Sect. 3 we obtain the thermodynamic properties which will be calculated using the expression for the partition function. In section 4, we calculate the effective Boltzmann factor $B(E)$ considering modified Dirac delta distribution in the deformed formalism. We obtain the statistical propertiesof the systems by using the superstatistics. In section 5, we obtain the rotational-vibrational energy spectrum for some diatomic molecules with numerical results and discussion. In section 6, we present special case of the potential under consideration. Finally, in section 7 we give concluding remark.

## 2. Energy levels and wavefunctions
The radial part of the Schrödinger equation is given by[60];

$$\frac{d^2 R_{n\ell}(r)}{dr^2} + \frac{2\mu}{\hbar^2}\left[E_{n\ell} - V(r) - \frac{\hbar^2 \ell(\ell+1)}{2\mu r^2}\right] R_{n\ell}(r) = 0 \qquad (2)$$

Considering the Generalized trigonometric Pöschl–Teller potential (Eq.(1b)), we obtain the radial Schrödinger equation, Eq.(2) is rewritten as follows:

$$\frac{d^2 R_{n\ell}(r)}{dr^2} + \frac{2\mu}{\hbar^2}\left[\begin{array}{c} E_{n\ell} - \left(V_1 \cos ec^2(\alpha r) + V_2 \sec^2(\alpha r) + V_3 \tan^2(\alpha r) + V_4 \cot^2(\alpha r)\right) \\ -\frac{\hbar^2 \ell(\ell+1)}{2\mu r^2} \end{array}\right] R_{n\ell}(r) = 0 \text{ S(3)}$$

This equation cannot be solved analytically for $\ell \neq 0$ due to the centrifugal term. Therefore, we must use an approximation to the centrifugal term. We use the following approximation[57]

$$\frac{1}{r^2} \cong \alpha^2 \left[d_0 + \frac{1}{\sin^2(\alpha r)}\right] \qquad (4)$$

where $d_0 = \frac{1}{12}$ is a dimensionless shifting parameter. This approximation scheme is an improved version of Greene and Aldrich[19, 53] approximation scheme.Inserting Eqs. (4) into Eq. (3), we have;

$$\frac{d^2 R_{n\ell}(r)}{dr^2} + \frac{2\mu}{\hbar^2}\left[ E_{n\ell} - \left(V_1 \csc^2(\alpha r) + V_2 \sec^2(\alpha r) + V_3 \tan^2(\alpha r) + V_4 \cot^2(\alpha r)\right) - \frac{\hbar^2 \ell(\ell+1)\alpha^2}{2\mu}\left(d_0 + \frac{1}{\sin^2(\alpha r)}\right) \right] R_{n\ell}(r) = 0 \quad (5)$$

Using the coordinate transformation $\rho = \sin^2(\alpha r)$, Eq.(5) translates into,

$$4\rho(1-\rho)\frac{d^2 R_{n\ell}(\rho)}{d\rho^2} + (2-4\rho)\frac{dR_{n\ell}(\rho)}{d\rho} + \frac{1}{\rho(1-\rho)}\left[\begin{array}{c}-(\varepsilon + \eta_3 + \eta_4 - \gamma d_0)\rho^2 \\ +(\varepsilon + \eta_1 - \eta_2 + 2\eta_4 - \gamma d_0 + \gamma)\rho - (\eta_1 + \eta_4 + \gamma)\end{array}\right] R_{n\ell}(\rho) = 0. \quad 6$$

For Mathematical simplicity, let's introduce the following dimensionless notations;

$$\varepsilon = \frac{2\mu E_{n\ell}}{\hbar^2 \alpha^2}, \eta_i = \frac{2\mu V_i}{\hbar^2 \alpha^2}, i=1,2,3,4, \gamma = \ell(\ell+1). \quad (7)$$

If we consider the following boundary conditions:

$$\rho \Rightarrow \begin{cases} 0, & r \to \infty, \\ 1, & r \to 0, \end{cases} \quad (8)$$

In view of the above boundary conditions, we propose the physical wave function as:

$$R_{n\ell}(\rho) = \rho^\beta (1-\rho)^\delta f(\rho) \quad (9)$$

where

$$\beta = \frac{1}{4} + \sqrt{\frac{1}{16} + \frac{(\eta_1 + \eta_4 + \gamma)}{4}} \quad (10)$$

$$\delta = \frac{1}{4} + \sqrt{\frac{1}{16} + \frac{(\eta_3 + \eta_2)}{4}} \quad (11)$$

On substitution of Eq. (9) into Eq. (6) leads to the following hypergeometric equation:

$$\rho(1-\rho)f''(\rho) + \left[\left(2\beta + \frac{1}{2}\right) - (2\beta + 2\delta + 1)\rho\right]f'(\rho) - \left[(\beta+\delta)^2 - \frac{(\varepsilon + \eta_3 + \eta_4 - \gamma d_0)}{4}\right]f(\rho) = 0 \quad (12)$$

whose solutions are the hypergeometric functions

$$f(\rho) = {}_2F_1(a,b;c;\rho) \quad (13)$$

where

$$a = (\beta+\delta) - \frac{\sqrt{\varepsilon + \eta_3 + \eta_4 - \gamma d_0}}{2}$$

$$b = (\beta+\delta) + \frac{\sqrt{\varepsilon + \eta_3 + \eta_4 - \gamma d_0}}{2} \quad (14)$$

$$c = 2\beta + \frac{1}{2}$$

By considering the finiteness of the solutions, the quantum condition is given by

$$(\beta+\delta) - \frac{\sqrt{\varepsilon + \eta_3 + \eta_4 - \gamma d_0}}{2} = -n \quad n=0,1,2... \quad (15)$$

from which we obtain, the energy expression as

$$\varepsilon = \gamma d_0 - \eta_3 - \eta_4 + \frac{1}{4}\left[4n+2+\sqrt{1+4(\eta_3+\eta_2)}+\sqrt{1+4(\eta_1+\eta_4+\gamma)}\right]^2 \tag{16}$$

Thus, if one substitutes the value of the dimensionless parameters in Eq.(7) into Eq.(16), we obtain the energy eigenvalues as:

$$E_{n\ell} = \frac{\hbar^2\alpha^2\ell(\ell+1)d_0}{2\mu} - V_3 - V_4 + \frac{\hbar^2\alpha^2}{8\mu}\left[4n+2+\sqrt{1+\frac{8\mu V_3}{\hbar^2\alpha^2}+\frac{8\mu V_2}{\hbar^2\alpha^2}}+\sqrt{\frac{8\mu V_1}{\hbar^2\alpha^2}+\frac{8\mu V_4}{\hbar^2\alpha^2}+(2\ell+1)^2}\right]^2 \tag{17}$$

The corresponding unnormalized wave function is obtain as

$$R_{n\ell}(\rho) = N_{n\ell}\rho^\beta(1-\rho)^\delta \,_2F_1\left(-n,n+2(\beta+\delta),2\beta+\frac{1}{2},\rho\right) \tag{18}$$

### 3. Thermal Properties of generalized trigonometric Pöschl–Teller potential.

We consider the contribution of the bound state to the rotational-vibrational partition function at a given temperature $T$ [58, 60]

$$Z(\beta) = \sum_{n=0}^{n_{max}} e^{-\beta E_{n\ell}}, \quad \beta = \frac{1}{k_B T} \tag{19}$$

Here, $k_B$ is the Boltzmann constant and $E_{n\ell}$ is the rotational-Vibrational energy of the nth bound state.

We can rewrite eq. (17) to be of the form

$$E_{n\ell} = \sigma_1 + \frac{\hbar^2\alpha^2}{8\mu}(4n+\sigma_2)^2 \tag{20}$$

where

$$\sigma_1 = \frac{\hbar^2\alpha^2\ell(\ell+1)d_0}{2\mu} - V_3 - V_4; \quad \sigma_2 = 2+\sqrt{1+\frac{8\mu V_3}{\hbar^2\alpha^2}+\frac{8\mu V_2}{\hbar^2\alpha^2}}+\sqrt{\frac{8\mu V_1}{\hbar^2\alpha^2}+\frac{8\mu V_4}{\hbar^2\alpha^2}+(2\ell+1)^2} \tag{21}$$

We substitute eq. (20) into eq. (19) to have

$$Z(\beta) = \sum_{n=0}^{n_{max}} e^{-\beta\left[\sigma_1+\frac{\hbar^2\alpha^2}{8\mu}(4n+\sigma_2)^2\right]} \tag{22}$$

where

$$n_{max} = \left|\frac{\sigma_2}{4}\right| \tag{23}$$

Replacing the sum in Eq.(22) by an integral in the classical limit, we obtain

$$Z(\beta) = \int_0^{n_{max}} e^{-\beta(An^2+Bn+C)} dn \qquad (24)$$

where

$$A = \frac{2\hbar^2\alpha^2}{\mu}; B = \frac{\hbar^2\alpha^2\sigma_2}{\mu}; C = \frac{\hbar^2\alpha^2\sigma_2^2}{8\mu} + \sigma_1 \qquad (25)$$

We therefore use the Maple software to evaluate the integral in eq. (24), thus obtaining the rotational-vibrational partition function with Generalised Trigonometric Posch-Teller potential models as

$$Z(\beta) = \frac{\sqrt{\pi} e^{\frac{\beta B^2}{4A}} \left( erf\left(\frac{\beta(2An_{max}+B)}{2\sqrt{\beta A}}\right) - erf\left(\frac{\beta B}{2\sqrt{\beta A}}\right) \right) e^{-\beta C}}{2\sqrt{\beta A}} \qquad (26)$$

The imaginary error function can be defined as [62]

$$erfi(z) = ierf(z) = \frac{2}{\sqrt{\pi}} \int_0^z e^{u^2} du \qquad (27)$$

Thermodynamic functions such as Helmholtz free energy, $F(\beta)$, entropy, $S(\beta)$, internal energy, $U(\beta)$, and specific heat, $C_v(\beta)$, functions can be obtained from the partition function(30) as follows [20];

$$F(\beta) = -\frac{1}{\beta} \ln Z(\beta) \qquad (28)$$

$$S(\beta) = -k_\beta \frac{\partial F(\beta)}{\partial \beta} \qquad (29)$$

$$U(\beta) = -\frac{\partial(\ln Z(\beta))}{\partial \beta} \qquad (30)$$

$$C_v = k_\beta \frac{\partial U(\beta)}{\partial \beta} \qquad (31)$$

### 4. Superstatistics Mechanics

In this section, we introduce the necessary conditions of superstatistics. The effective Boltzmann factor of the system can be written as[73, 74]

$$B(E) = \int_0^\infty e^{-\beta' E} f(\beta',\beta) d\beta' \qquad (32)$$

$$f(\beta',\beta) = \delta(\beta'-\beta) \tag{33}$$

Finally, we find the generalized Boltzmann factor[75]

$$B(E) = e^{-\beta E}\left(1 + \frac{q}{2}\beta^2 E^2\right) \tag{34}$$

where $q$ is the deformation parameter. Details of Eq. (34) can be found in Appendix A of ref[75] and references therein.

The partition function for the modified Dirac delta distribution has the following form:

$$Z_S = \int_0^\infty B(E)dn \tag{35}$$

We substitute eq. (20) into eq. (35) to have

$$Z_S = \int_0^\infty e^{-\beta\left(\sigma_1 + \frac{\hbar^2\alpha^2}{8\mu}(4n+\sigma_2)^2\right)}\left(1 + \frac{q}{2}\beta^2\left(\sigma_1 + \frac{\hbar^2\alpha^2}{8\mu}(4n+\sigma_2)^2\right)^2\right)dn \tag{36}$$

We therefore use the Mathematica software to evaluate the integral in eq. (36), thus obtaining the partition function with Generalized Trigonometric Posch-Teller potential model in superstatistics as follows

$$Z_S(\beta) = \frac{e^{-\beta(\sigma_1+\chi\sigma_2^2)}\left(e^{\beta\chi\sigma_2^2}\sqrt{\pi}\left(8+3q+4q\beta\sigma_1(1+\beta\sigma_1)\right) - \sqrt{\beta\chi}(\xi)\sigma_2 + 4q(\beta\chi)^{\frac{3}{2}}\sigma_2^3\right)}{64\sqrt{\beta\chi}} \tag{37}$$

where

$$\xi = -6q - 8q\beta\sigma_1 + \frac{e^{\beta\chi\sigma_2^2}\sqrt{\pi}\,\text{Erf}\left[\sqrt{\beta}\sqrt{\chi}\sigma_2\right]}{\sqrt{\beta}\sqrt{\chi}\sigma_2} \tag{38}$$

Other thermodynamic functions such as Helmholtz free energy, $F_S(\beta)$, entropy, $S_S(\beta)$, internal energy, $U_S(\beta)$, and specific heat, $C_S(\beta)$, functions can be obtained from the partition function(38) with the aid of eqs.(28-31).

## 5. Numerical Results and Applications

To show the accuracy of our work, we calculate the energy eigenvalues using Eq.(18) for different quantum numbers $n$ and $\ell$ with parameters $V_1 = 5\,fm^{-1}$, $V_2 = 3\,fm^{-1}$, $V_3 = 0.5$, $V_4 = 0.5$ and $\mu = 10\,fm^{-1}$. In Table 1., it is observed that the energy decreases for a fixed value of the principal quantum number for varying orbital angular momentum. Furthermore, we have computed the energy eigenvalues of the Trigonometric Posch-Teller potential using the reduced energy equation given in Eq. (33) and Eq.(34) as special case. Our results shown in Tables 2- 5 are in good agreement with the results given in Ref. [56-57,76]. Fig. 1 shows a comparative plot of the shapes of the trigonometric Poschl-Teller potential model and generalized trigonometric Posch-Teller potential. In Fig. 2, we plot the shape of the generalized trigonometric Posch-Teller potential for different values of the screening parameter $\alpha$. Figs. 3-6 clearly shows the energy eigenvalues variation with parameters $V_1, V_2, V_3$ and $V_4$ for various quantum states. It can be easily observed from these Figs. 3-6 that the parameters increases directly as the energy increases. Fig.7 shows the energy eigenvalues variation with the particle's reduced mass $\mu$ for different quantum states. It is seen that in the region $\mu \approx 0-0.1\, a.m.u$, the energy eigenvalue is at its maximum, beyond this region, there is a drop and this continues in a linear trend. The energy is only high in the region where the mass is low but decreases as the particle's mass increases monotonically. The energy is very similar for $0.2 < \mu < 1.0$. Fig. 8 shows the energy eigenvalues variation with orbital quantum number for various principal quantum number. It is shown in the plot that the energy increases as the principal quantum number increases. Fig. 9 shows the energy eigenvalues variation with screening parameter $\alpha$ for different quantum states, it can be seen explicitly that in all the quantum states the representation curves spreads out uniformly from the origin. It is shown that the energy eigenvalue increases as the screening parameter increases. Fig. 10 shows the vibrational partition function variation with $\beta$ for various values of $\lambda$. It can be seen that the partition function decreases as the temperature increases. It is also shown that in the high temperature, there's a uniform convergence of all the curves and the partition function reaches its minimum. Fig. 11 shows vibrational partition function variation with $\lambda$ for various values of $\beta$. From this plot, we observe that there is spread from the zero point, the partition function increases as $\lambda$ increases. It is also observed that the partition function has its maximum in the high region of $\lambda$. Fig. 12 shows the mean vibrational energy variation with $\beta$ for various values of $\lambda$. The mean vibrational energy decreases as $\beta$ increases. It is also observed that the mean vibrational energy has its minimum in the high $\beta$ region. Fig. 13 clearly shows the vibrational specific heat capacity variation with $\beta$ for different values of $\lambda$. It can be seen that the specific heat capacity increases as $\beta$ increases. Fig. 14 shows the vibrational entropy variation with $\beta$ for different values of $\lambda$. It is seen that the vibrational entropy decreases as temperature increases. Fig. 15 shows the vibrational entropy variation with $\lambda$ for different values of $\beta$. It is seen that the entropy increases monotonically with $\lambda$. Fig. 16 shows the mean vibrational free energy variation with $\beta$ for different values of $\lambda$. Again, it is seen that the vibrational free energy increases monotonically with $\beta$. Fig. 17 shows the mean vibrational free energy variation with $\lambda$ for different values of $\beta$. It is clearly shown that the mean vibrational free energy decreases as $\lambda$ increases. In Fig. 18-22, we study the statistical properties of the system by using the superstatistics formalism. Fig. 18 shows a plot of the vibrational partition function variation with $\beta$ for various values of $q$. It is seen that the partition function decreases as $\beta$ increases. More so, the partition function increases as q increases. Fig. 19 shows the variation of vibrational mean free energy with $\beta$ for various values of $q$. The mean free energy decreases monotonically as $\beta$ increases. A plot of vibrational entropy variation with $\beta$ for different values of $q$ is shown in Fig. 20. The entropy of the system reduces as $\beta$ rises. For different values of the deformation parameter, $q$, the entropy of the system increases with increasing $q$. Fig. 21 shows the variation of the vibrational mean energy with $\beta$ for various values of $q$. The mean energy decreases with increasing $\beta$ and increases with increasing $q$. Fig. 22 depicts a plot of vibrational specific heat capacity variation with $\beta$ for different values of $q$. It is seen explicitly that the specific heat capacity of the system increases monotonically with increasing $\beta$ but decreases with increasing $q$. It is interesting to note also that when $q=0$, normal statistics is recovered.

## 6. Special Cases

In this section, we shall study one special case of the Generalized Trigonometric Posch-Teller potential and its energy eigenvalues respectively

### 6.1 Trigonometric Posch-Teller potential

Choosing $V_3 = V_4 = 0$, the Generalized Trigonometric Posch-Teller potential takes [76,56]

$$V(r) = \frac{V_1}{\sin^2(\alpha r)} + \frac{V_2}{\cos^2(\alpha r)} \tag{39}$$

and the energy eigenvalue becomes

$$E_{n\ell} = \frac{\hbar^2 \alpha^2 \ell(\ell+1)d_0}{2\mu} + \frac{\hbar^2 \alpha^2}{8\mu}\left[4n + 2 + \sqrt{1 + \frac{8\mu V_2}{\hbar^2 \alpha^2}} + \sqrt{\frac{8\mu V_1}{\hbar^2 \alpha^2} + (2\ell+1)^2}\right]^2 \tag{40}$$

This is in excellent agreement with Eq. (3) of Ref.[76] and Eq.(19) of Ref.[57].

By setting $\ell = 0$, we obtain the s-wave energy equation for the potential understudy as;

$$E_{n0} = \frac{\hbar^2 \alpha^2}{8\mu}\left[4n + 2 + \sqrt{1 + \frac{8\mu V_2}{\hbar^2 \alpha^2}} + \sqrt{\frac{8\mu V_1}{\hbar^2 \alpha^2} + 1}\right]^2 \tag{41}$$

This is in excellent agreement with Eq. (15) of Ref. [56].

## 7. Conclusion
In this article, we have solved the Schrodinger equation using Functional Analysis Approach and suitable approximation to overcome the centrifugal barrier. We have also presented the rotational-vibrational energy spectra with the Generalized Trigonometric Posch-Teller potential. We have expressed the solutions by the generalized hypergeometric functions $_2F_1(a,b;c;\rho)$. Results have been discussed extensively using plots. We discussed some special cases by adjusting the potential parameters and compute the numerical energy spectra for the Trigonometric Posch-Teller potential for both the $\ell = 0$ and $\ell \neq 0$ cases respectively. It was found that our results agree with the existing literature. In detail, we evaluated the vibrational partition functions $Z(\beta)$ which we used to study the thermodynamics properties of vibrational mean energy $U(\beta)$, vibrational entropy $S(\beta)$, vibrational mean free energy $F(\beta)$ and vibrational specific heat capacity $C(\beta)$. In addition, the effective Boltzmann factor is calculated by using superstatistics and the results is compared with the case of where the deformation parameter vanished. It is noted that the results, in the special case of the vanished deformation parameter, are in agreement with the ordinary statistics.Finally, this study has many applications in different areas of physics and chemistry such as atomic physics, molecular physics and chemistry amongst others

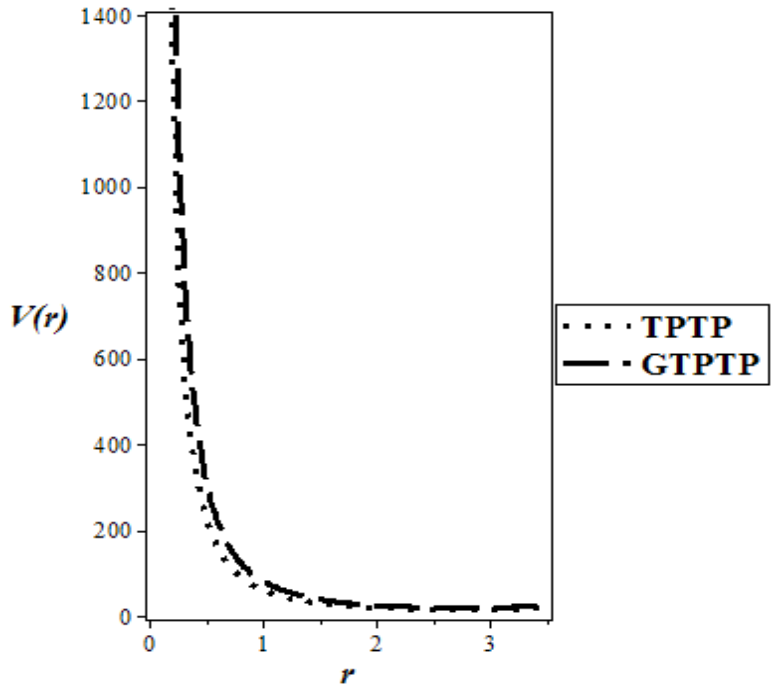

**Fig. 1**: Shape of the trigonometric Poschl– Teller potential model andgeneralized trigonometric Posch-Teller potential. We chose $V_1 = 5\,fm^{-1}$, $V_2 = 3\,fm^{-1}$, $V_3 = 2\,fm^{-1}$, $V_4 = 0.5\,fm^{-1}$ and $\alpha = 0.3$

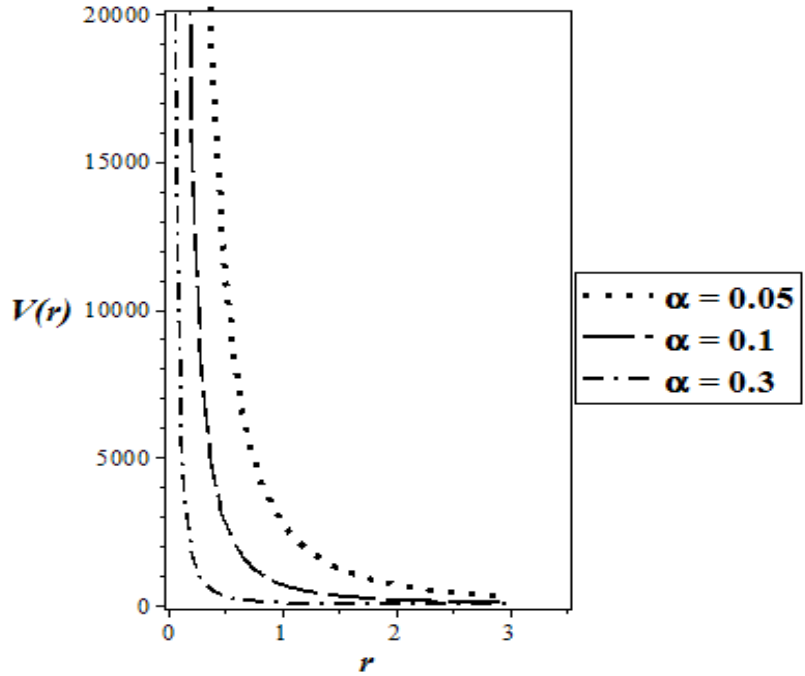

**Fig. 2**: Shape of the generalized trigonometric Posch-Teller potential for different values of the screening parameter $\alpha$. We chose $V_1 = 5\,fm^{-1}$, $V_2 = 3\,fm^{-1}$, $V_3 = 2\,fm^{-1}$ and $V_4 = 0.5\,fm^{-1}$

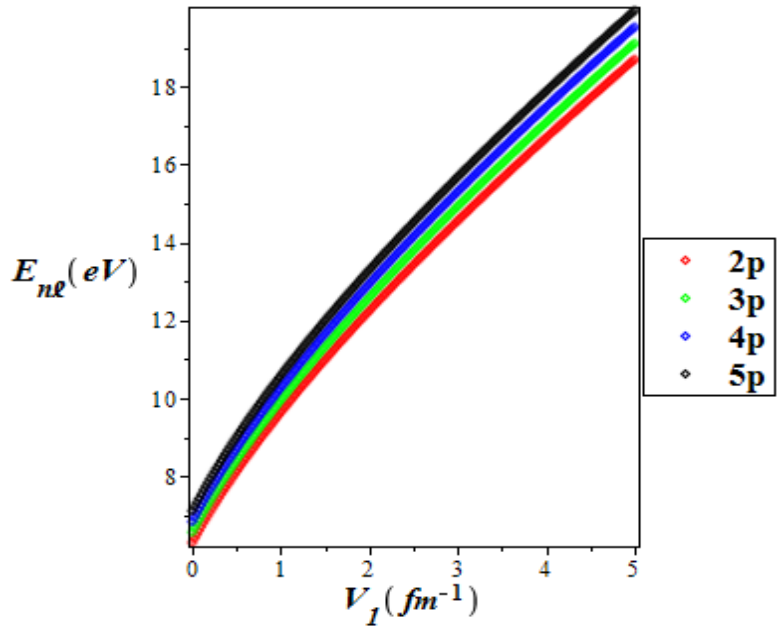

**Fig. 3**: Energy eigenvalues variation with parameter $V_1$ for various quantum states

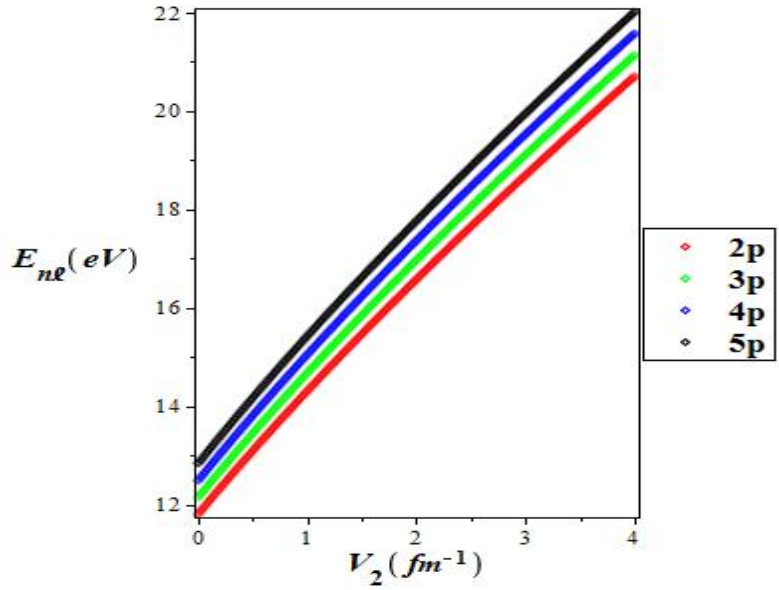

**Fig. 4**: Energy eigenvalues variation with parameter $V_2$ for various quantum states

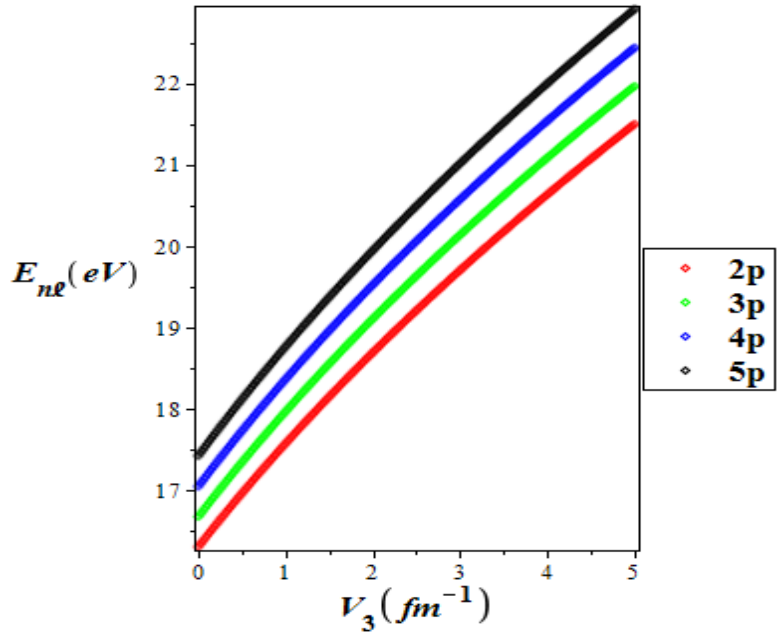

**Fig. 5**: Energy eigenvalues variation with parameter $V_3$ for various quantum states

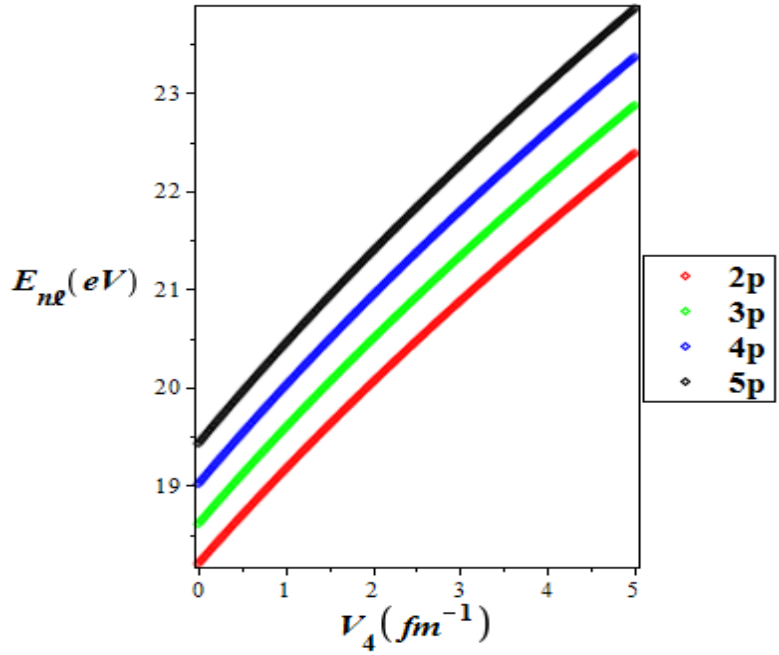

**Fig. 6**: Energy eigenvalues variation with parameter $V_4$ for various quantum states

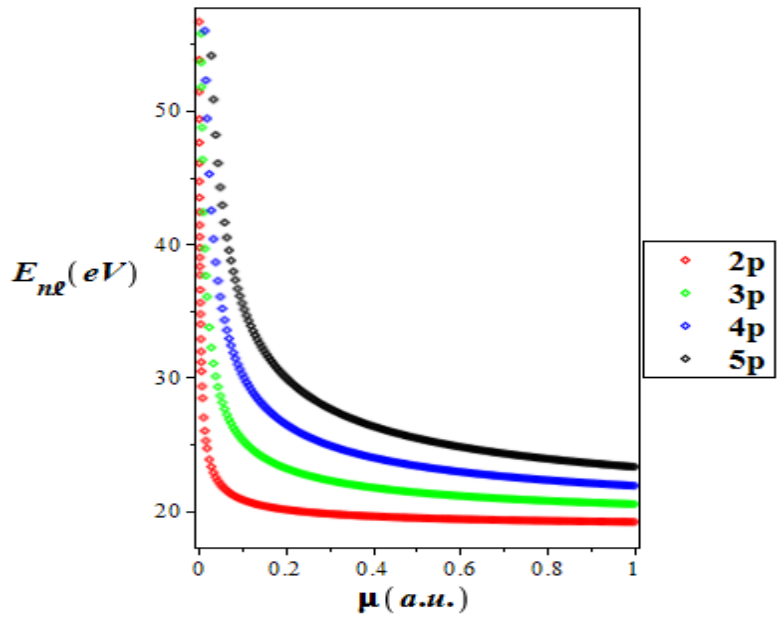

**Fig. 7**: Energy eigenvalues variation with particle's mass $\mu$ for various quantum states

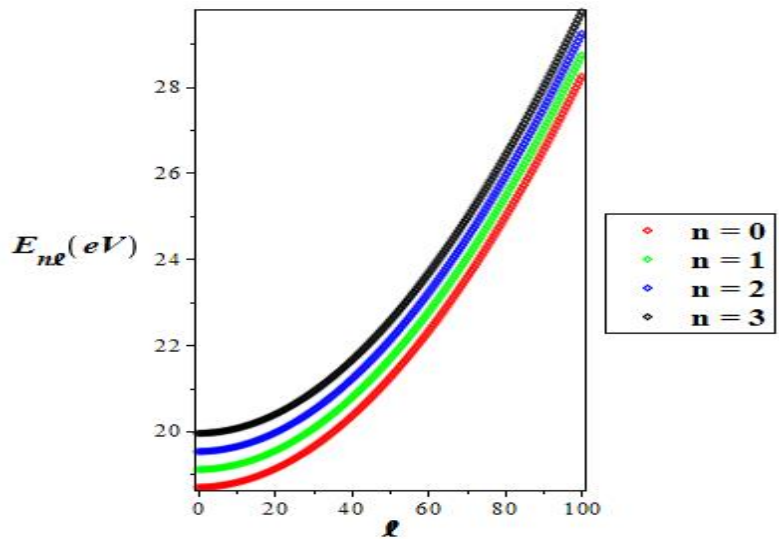

**Fig. 8**: Energy eigenvalues variation with rotational quantum number for various vibrational quantum numbers.

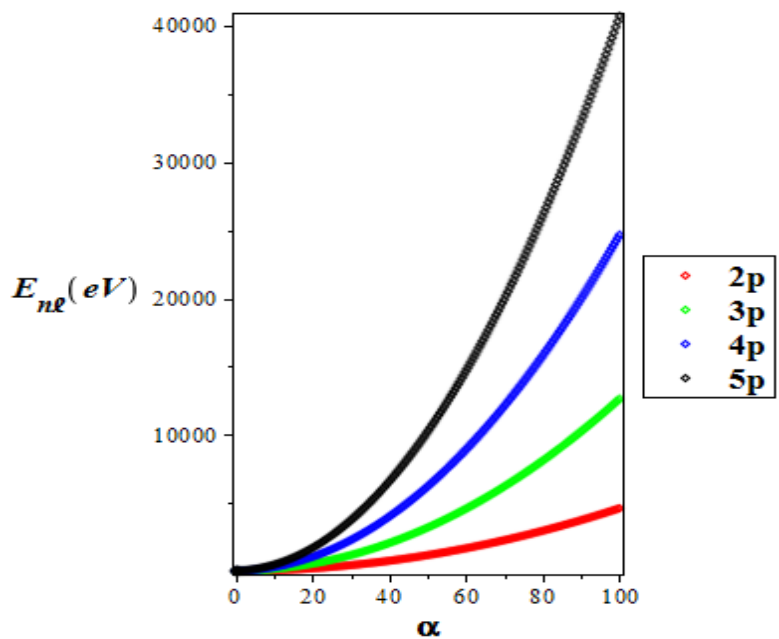

**Fig. 9**: Energy eigenvalues variation with screening parameter $\alpha$ for quantum states

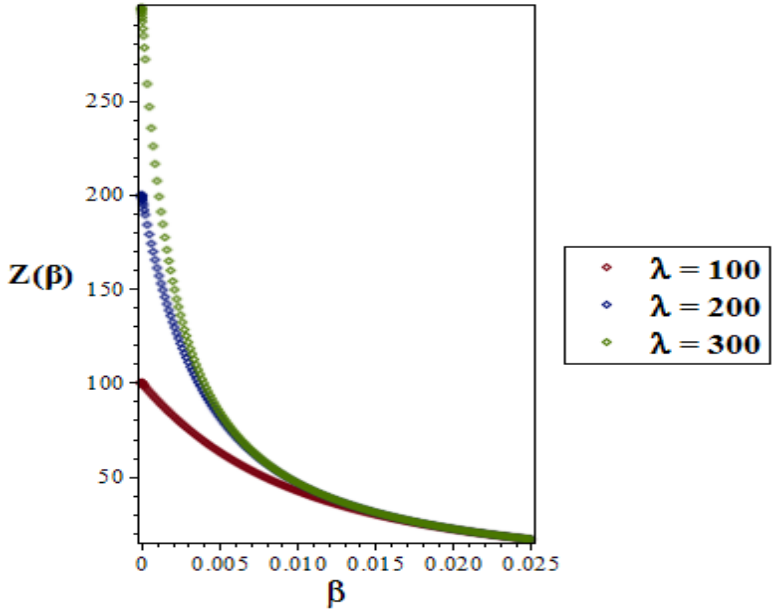

**Fig. 10**: Vibrational Partition Function variation with $\beta$ for various values of $\lambda$

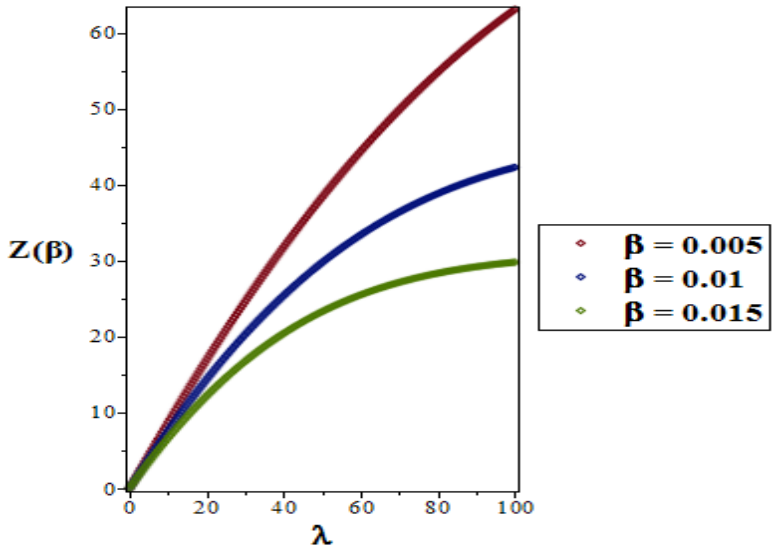

**Fig. 11**: Vibrational Partition Function variation with $\lambda$ for various values of $\beta$

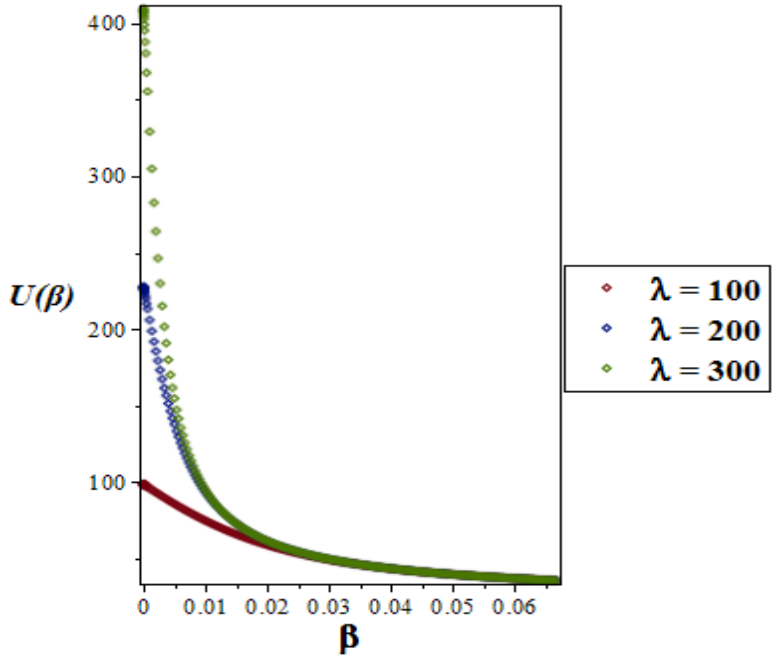

**Fig. 12**: Vibrational mean energy variation with $\beta$ for various values of $\lambda$

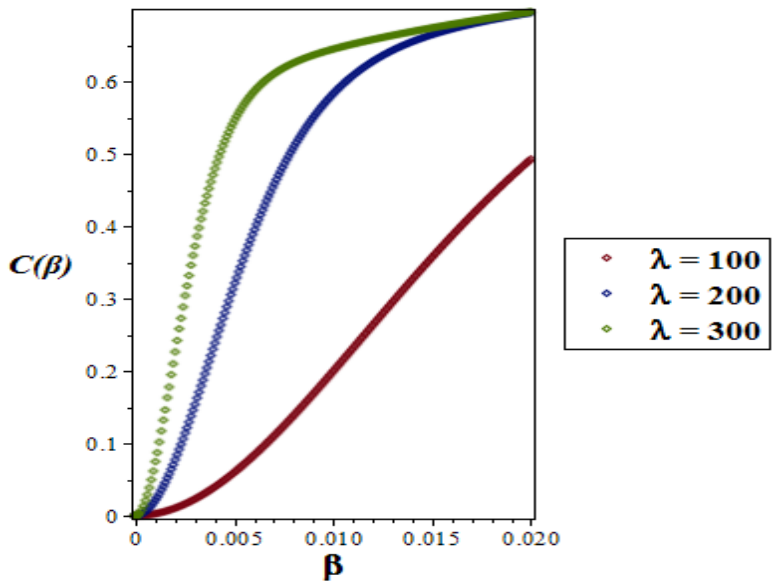

**Fig. 13**: Vibrational specific heat capacity variation with $\beta$ for different values of $\lambda$

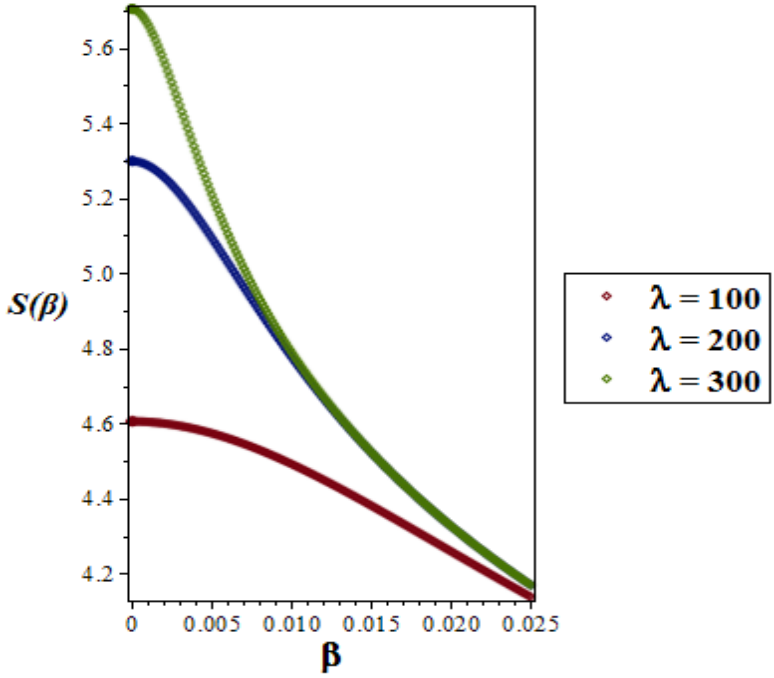

**Fig. 14**: Vibrational entropy variation with $\beta$ for different values of $\lambda$

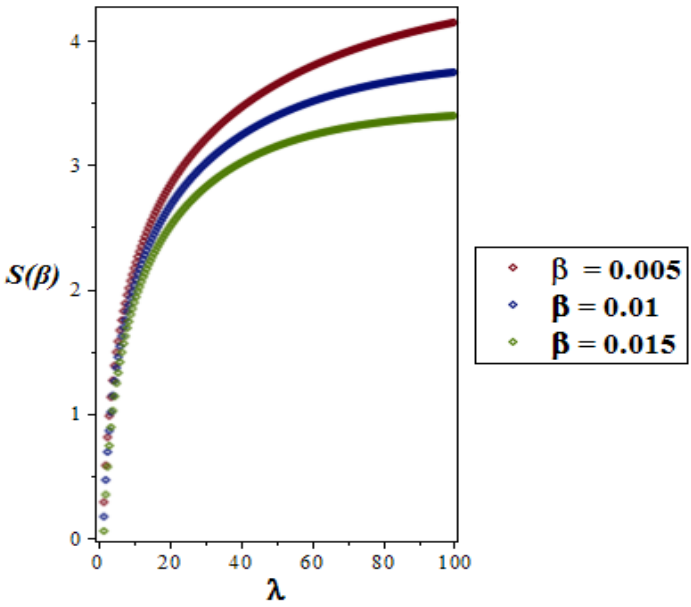

**Fig. 15**: Vibrational entropy variation with $\lambda$ for different values of $\beta$

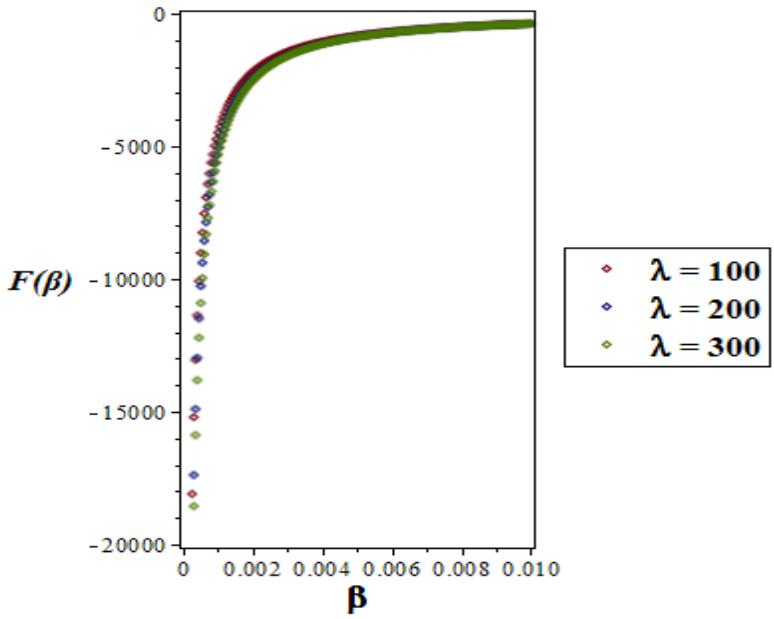

**Fig. 16**: Vibrational mean free energy variation with $\beta$ for different values of $\lambda$

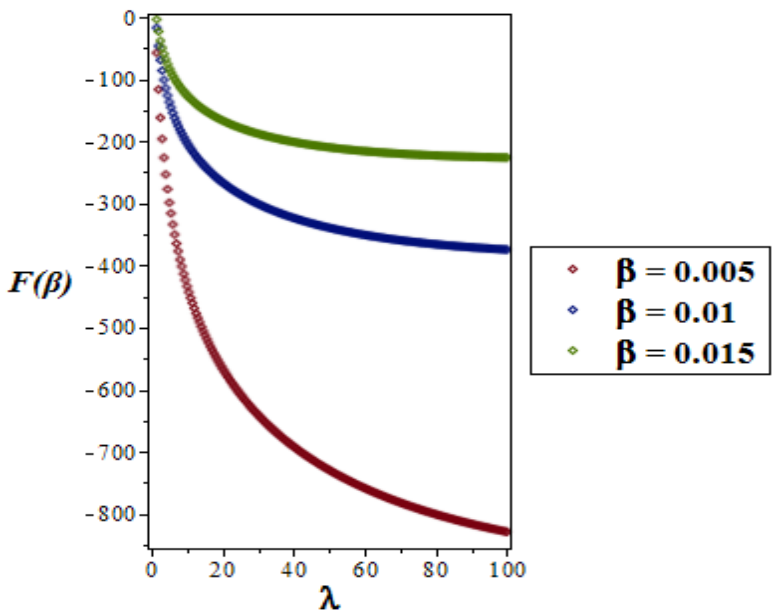

**Fig. 17**: Vibrational mean free energy variation with $\lambda$ for different values of $\beta$

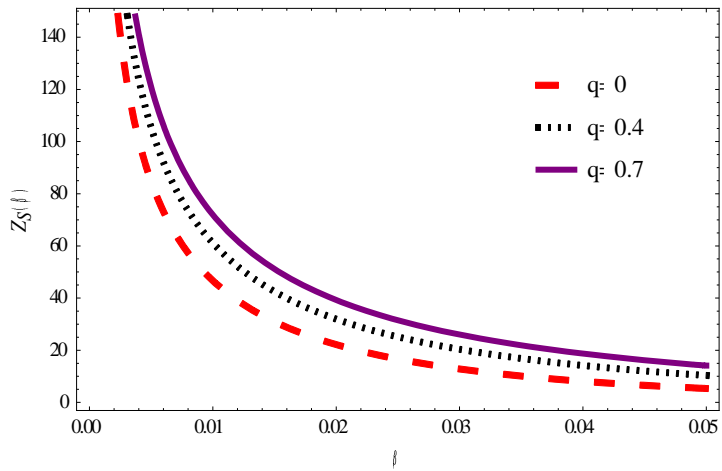

**Fig. 18**: Vibrational Partition Function variation with $\beta$ for various values of $q$

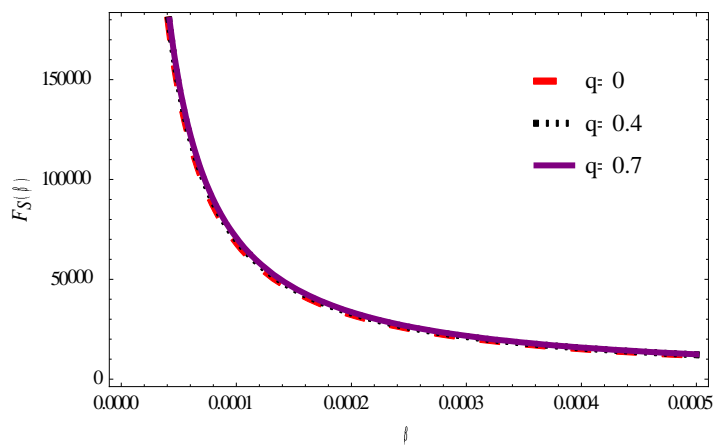

**Fig. 19**: Vibrational meanfree energy variation with $\beta$ for various values of $q$

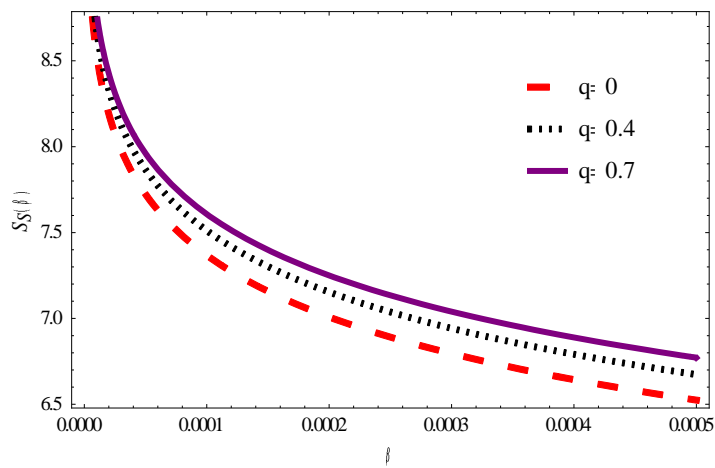

**Fig. 20**: Vibrational entropy variation with $\beta$ for different values of $q$

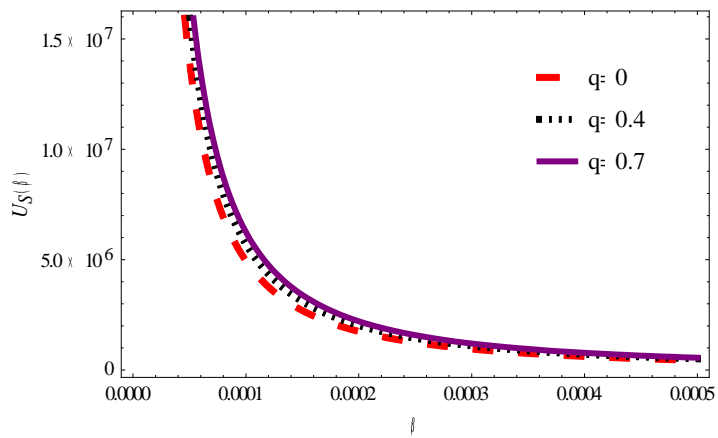

**Fig. 21**: Vibrational mean energy variation with $\beta$ for various values of $q$

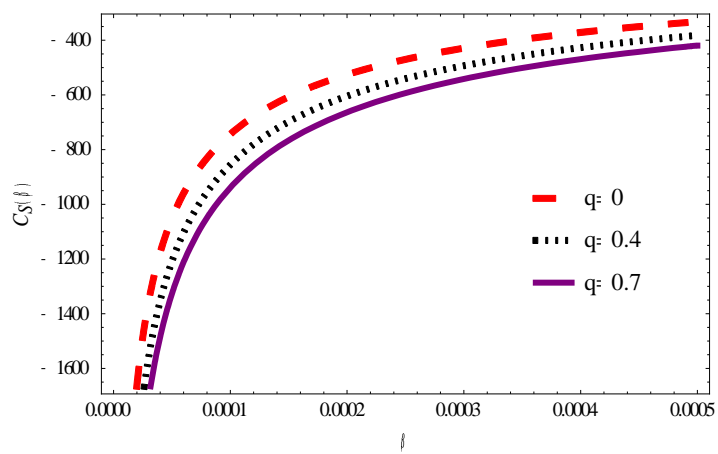

**Fig. 22**: Vibrational specific heat capacity variation with $\beta$ for different values of $q$

**Table 1**: Bound state energy levels $E_{n\ell}$ for the Generalised trigonometric Posch-Teller potential obtained with parameters $V_1 = 5\,fm^{-1}$, $V_2 = 3\,fm^{-1}$, $V_3 = 0.5$, $V_4 = 0.5$ and $\mu = 10\,fm^{-1}$.

| States | $\alpha = 0.002$ | $\alpha = 0.02$ | $\alpha = 0.2$ | $\alpha = 0.4$ | $\alpha = 0.8$ | $\alpha = 1.2$ |
|---|---|---|---|---|---|---|
| 1s | 18.50038334 | 18.61121709 | 19.73743352 | 21.02713244 | 23.72899708 | 26.59648864 |
| 2s | 18.50858178 | 18.69348948 | 20.58899614 | 22.79449204 | 27.52159072 | 32.67406160 |
| 2p | 18.50858258 | 18.69357133 | 20.59752395 | 22.83011829 | 27.67609429 | 33.04821322 |
| 3s | 18.51678180 | 18.77592188 | 21.45655875 | 24.62585164 | 31.57018435 | 39.32763454 |
| 3p | 18.51678262 | 18.77600388 | 21.46523908 | 24.66269723 | 31.73441612 | 39.73447216 |
| 3d | 18.51678425 | 18.77616788 | 21.48259499 | 24.73630368 | 32.06125854 | 40.53875474 |
| 4s | 18.52498345 | 18.85851426 | 22.34012135 | 26.52121126 | 35.87477799 | 46.55720749 |
| 4p | 18.52498425 | 18.85859642 | 22.34895421 | 26.55927615 | 36.04873794 | 46.99673112 |
| 4d | 18.52498588 | 18.85876072 | 22.36661499 | 26.63531596 | 36.39487055 | 47.86516178 |
| 4f | 18.52498834 | 18.85900721 | 22.39309395 | 26.7491527 | 36.90974960 | 49.14313806 |

**Table 2**; Comparison of s-wave energy eigenvalues (in eV) obtained by using the Functional Analysis Approach with other methods for the trigonometric Poschl–Teller potential with other methods obtained with parameters $V_1 = 5\,fm^{-1}$, $V_2 = 3\,fm^{-1}$, and $\mu = 10\,fm^{-1}$.

| n | Present | AIM[64], $\alpha = 0.2$ | NU[56] $\alpha = 0.2$ | Present | AIM[64] $\alpha = 0.02$ | NU[56] $\alpha = 0.02$ | Present | AIM[64] $\alpha = 0.002$ | NU[56] $\alpha = 0.002$ |
|---|---|---|---|---|---|---|---|---|---|
| 0 | 16.10494172 | 16.104 941 73 | 16.104 941 72 | 15.78149898 | 15.781 498 98 | 15.781 498 98 | 15.7495163 | 15.749 516 29 | 15.749 516 29 |
| 1 | 16.83082621 | 16.830 826 21 | 16.830 826 21 | 15.8526429 | 15.852 642 89 | 15.852 642 89 | 15.75661628 | 15.756 616 28 | 15.756 616 28 |
| 2 | 17.5727107 | 17.572 710 70 | 17.572 710 70 | 15.9239468 | 15.923 946 80 | 15.923 946 80 | 15.76371788 | 15.763 717 86 | 15.763 717 86 |
| 3 | 18.33059519 | 18.330 595 18 | 18.330 595 18 | 15.99541072 | 15.995 410 71 | 15.995 410 71 | 15.77082105 | 15.770 821 05 | 15.770 821 05 |
| 4 | 19.10447968 | 19.104 479 67 | 19.104 479 67 | 16.06703462 | 16.067 034 63 | 16.067 034 63 | 15.77792584 | 15.777 925 84 | 15.777 925 84 |
| 5 | 19.89436415 | 19.894 364 16 | 19.894 364 16 | 16.13881855 | 16.138 818 54 | 16.138 818 54 | 15.78503222 | 15.785 032 22 | 15.785 032 22 |
| 6 | 20.70024864 | 20.700 248 64 | 20.700 248 64 | 16.21076245 | 16.210 762 45 | 16.210 762 45 | 15.79214021 | 15.792 140 21 | 15.792 140 21 |

**Table 3**; Comparison of s-wave energy eigenvalues (in eV) obtained by using the Functional Analysis Approach with other methods for the trigonometric Poschl– Teller potential with other methods obtained with parameters $V_1 = 5\,fm^{-1}$, $V_2 = 3\,fm^{-1}$, and $\mu = 10\,fm^{-1}$.

| n | Present | NU[56] $\alpha=1.2$ | Present | NU[56] $\alpha=0.8$ | Present | NU[56] $\alpha=0.4$ |
|---|---|---|---|---|---|---|
| 0 | 18.02560022 | 18.02560022 | 17.23163309 | 17.23163309 | 16.47211972 | 16.47211973 |
| 1 | 22.87051711 | 22.8705171 | 20.32991862 | 20.32991862 | 17.95616358 | 17.95616357 |
| 2 | 28.29143400 | 28.29143398 | 23.68420415 | 23.68420415 | 19.50420741 | 19.50420742 |
| 3 | 34.28835088 | 34.28835086 | 27.29448969 | 27.2944896 | 21.11625128 | 21.11625126 |
| 4 | 40.86126776 | 40.86126774 | 31.16077521 | 31.16077522 | 22.79229512 | 22.7922951 |
| 5 | 48.01018464 | 48.01018462 | 35.28306075 | 35.28306074 | 24.53233896 | 24.53233894 |
| 6 | 55.73510152 | 55.7351015 | 39.66134628 | 39.66134628 | 26.33638282 | 26.33638278 |

**Table 4**; Comparison of l-state energy eigenvalues (in eV) obtained by using the Functional Analysis Approach with other methods for the trigonometric Poschl– Teller potential with other methods obtained with parameters $V_1 = 5\,fm^{-1}$, $V_2 = 3\,fm^{-1}$, and $\mu = 10\,fm^{-1}$.

| States | Present | NU[57] $\alpha=1.2$ | Present | NU[57] $\alpha=0.8$ | Present | NU[57] $\alpha=0.4$ | Present |
|---|---|---|---|---|---|---|---|
| 1s | 22.87051711 | 22.87051710 | 20.32991862 | 20.32991862 | 17.95616358 | 17.95616357 | 16.83082621 |
| 2s | 28.29143400 | 28.29143398 | 23.68420415 | 23.68420415 | 19.50420741 | 19.50420742 | 17.57271070 |
| 2p | 28.64395420 | 28.64395419 | 23.82847893 | 23.82847894 | 19.53712285 | 19.53712286 | 17.58054181 |
| 3s | 34.28835088 | 34.28835086 | 27.29448969 | 27.2944896 | 21.11625128 | 21.11625126 | 18.33059519 |
| 3p | 34.67512504 | 34.67512504 | 27.44896379 | 27.44896381 | 21.15044541 | 21.15044543 | 18.33858624 |
| 3d | 35.43921159 | 35.43921159 | 27.75631555 | 27.75631556 | 21.21875332 | 21.2187533 | 18.35456400 |
| 4s | 40.86126776 | 40.86126774 | 31.16077521 | 31.16077522 | 22.79229512 | 22.7922951 | 19.10447968 |
| 4p | 41.28229588 | 41.28229584 | 31.32544867 | 31.32544868 | 22.82776799 | 22.8277680 | 19.11263069 |
| 4d | 42.11348591 | 42.11348590 | 31.65300784 | 31.65300783 | 22.89862722 | 22.89862721 | 19.12892818 |
| 4f | 43.33519178 | 43.33519178 | 32.14003976 | 32.14003977 | 23.00470172 | 23.00470171 | 19.15336296 |

**Table 5**; Comparison of l-state energy eigenvalues (in eV) obtained by using the Functional Analysis Approach with other methods for the trigonometric Poschl– Teller potential with other methods obtained with parameters $V_1 = 5\,fm^{-1}$, $V_2 = 3\,fm^{-1}$, and $\mu = 10\,fm^{-1}$.

| States | Present | NU[57], $\alpha = 0.2$ | Present | NU[57] $\alpha = 0.02$ | Present | NU[57] $\alpha = 0.002$ |
|---|---|---|---|---|---|---|
| 1s | 16.83082621 | 16.83082621 | 15.8526429 | 15.85264289 | 15.75661628 | 15.75661628 |
| 2s | 17.5727107 | 17.5727107 | 15.9239468 | 15.9239468 | 15.76371788 | 15.76371786 |
| 2p | 17.58054181 | 17.58054181 | 15.92402152 | 15.92402153 | 15.76371861 | 15.76371860 |
| 3s | 18.33059519 | 18.33059518 | 15.99541072 | 15.99541071 | 15.77082105 | 15.77082105 |
| 3p | 18.33858624 | 18.33858626 | 15.99548559 | 15.9954856 | 15.77082179 | 15.77082179 |
| 3d | 18.35456400 | 18.35456399 | 15.99563535 | 15.99563534 | 15.77082329 | 15.77082328 |
| 4s | 19.10447968 | 19.10447967 | 16.06703462 | 16.06703463 | 15.77792584 | 15.77792584 |
| 4p | 19.11263069 | 19.1126307 | 16.06710967 | 16.06710967 | 15.77792658 | 15.77792658 |
| 4d | 19.12892818 | 19.12892817 | 16.06725975 | 16.06725974 | 15.77792808 | 15.77792806 |
| 4f | 19.15336296 | 19.15336297 | 16.06748486 | 16.06748485 | 15.77793030 | 15.77793030 |